\documentclass{emulateapj}
\newcommand{\be}{\begin{equation}}
\newcommand{\ee}{\end{equation}}
\newcommand{\bea}{\begin{eqnarray}}
\newcommand{\eea}{\end{eqnarray}}
\shortauthors{}
\begin{document}
\title{Particle Acceleration by Fast Modes in Solar Flares} 

\author{ Huirong Yan\altaffilmark{1}, A. Lazarian\altaffilmark{2}, and Vah\'e Petrosian\altaffilmark{3}}

\altaffiltext{1}{CITA, Toronto, ON M5S3H8, Canada, yanhr@cita.utoronto.ca }
\altaffiltext{2}{Department of Astronomy, University of Wisconsin, Madison, WI 53706, alazarian@wisc.edu}
\altaffiltext{3}{Department of Applied Physics, Stanford
University, Stanford, CA 94305, vahep@stanford.edu}

\begin{abstract}
We address the problem of particle acceleration in
solar flares by fast modes which may be excited during the reconnection and
undergo cascade and are subjected to damping.  We extend the calculations
beyond quasilinear approximation and compare the acceleration and scattering by
transit time damping and gyroresonance interactions. We find that the
acceleration is dominated by the so called transit time damping mechanism. We estimate
the total energy transferred into particles, and show that our approach provides
sufficiently accurate results
We compare this rate  with energy loss rate. Scattering by fast modes appears to
be sufficient to prevent the protons from escaping the system during the
acceleration. Confinement of electrons, on the other hand, requires the
existence of plasma waves. Electrons can be accelerated to GeV energies through
the process described here for solar flare conditions.
\end{abstract}


\section{Introduction}
The mechanism of energy  release and the process of its transfer to heating
and
acceleration of nonthermal particles in many magnetized astrophysical
plasmas in general, and solar
flares in particular, are still matter of considerable debate. Recent research
shows turbulence
may play an essential role in these processes. In the case of solar flares, it
is believed that the
energy comes  from release of stored magnetic energy via reconnection (see
Priest \& Forbes 2000). Turbulence is expected to develop since both ordinary
and magnetic Reynolds number are very large. Both observational evidence and
theoretical arguments suggest most of the energy is dissipated through
turbulence (Biskamp 2000; Lazarian \& Vishniac 1999, Lazarian et al. 2004; Shay et al. 2004). Recent
high resolution observations of solar flares by {\it Yohkoh} and {\it RHESSI}
satellites have
supplied ample evidence that, at least from the point of view of particle
acceleration, plasma
turbulence and plasma waves appear to be the most promising agent  not only for
the
acceleration mechanism but also the general energizing of flare plasma (see {\it
e.g.} Petrosian 2007 and Petrosian \& Liu
2004 and references cited there). This may also be true in other situations
(Lazarian et al. 2002, Liu,
Petrosian \& Melia 2004). A substantial progress in
understanding of incompressible (Shebalin, Matthaeus \& Montgomery 1983; Higdon
1983; Goldreich \& Sridhar 1995, hereafter GS95; Matthaeus et al 1996, Cho,
Lazarian \& Vishniac 2003; Biskamp 2003) and
compressible MHD turbulence (see Lithwick \& Goldreich 2001; Cho \& Lazarian
2002, 2003), as well as MHD-turbulence-particle interactions
(Chandran 2000, Yan \& Lazarian 2002, 2004, 2008, henceforth YL02, YL04, YL08,
respectively) has clarified many aspects of this problem.

It was demonstrated that the often
adopted Alfv\'en modes undergo anisotropy cascade mainly in  the direction
perpendicular to the underlying magnetic field $B_0$ with a Kolmogorov spectrum. As
a result they are inefficient in acceleration of particles. Slow modes which mix
with Alfv\'enic modes are also negligible for the same reason. Fast modes, on
the other hand,
develop on their own, as their phase
velocity is only marginally affected by mixing motions induced
by Alfv\'en modes.   According
to Cho \& Lazarian (2002) fast modes
follow an {\it isotropic} ``acoustic"
cascade\footnote{We discuss in \S 5 the limitations of the model of turbulence in Cho \& Lazarian (2002, 2003). On the basis of weak-turbulence theory, Chandran
(2005) has argued that high-frequency fast waves with $k_\|>>k_\bot$ interacting with other fast waves generate high-frequency Alfven-waves with $k_\|>>k_\bot$. We expect that the scattering by thus generated Alfven  modes to be similar to scattering by fast modes that created them. Therefore,
within the simplified approach adopted in the paper, we do not consider this type of interactions.} where the velocity scales as  $v_k\propto k^{-1/4}$ and in each
wave-wave collision a small fraction
of energy equal to
$v_{ph}/v_k$ is transferred to smaller scales. In low $\beta$ medium, the three dimensional
energy spectrum is (YL02)
\bea
W(k)= \frac{nm_i\delta V^2}{8\pi}k^{-\frac{7}{2}}L^{-\frac{1}{2}}\left(\begin{array}{cc}k_ik_j/k_\bot^2&\\ &0\end{array}\right),
\label{fastspec}
\eea
with a cascade time scale of
\be
\tau_{cas}=(v_{ph}/v_k)(kv_{k})^{-1}=(L/\delta V)M_A^{-1}(kL)^{-1/2}.
\label{tcasfast}
\ee
Here $k_{i,j}$ refers to the x,y components of the wave vector ${\bf k}$, $n$ is the density of the plasma with ions of mass $m_i$ ($\sim$ proton
mass), $\delta V$ is the initial perturbation injected at the outer scale $L\equiv
k_{\rm min}$), $v_{ph}=\omega/k$ is the phase speed of fast
mode with frequency $\omega$ and wave vector $k$, and $M_A=\delta V/v_A$ is the
Alfv\'enic Mach number for the Alfv\'en speed $\simeq v_A=B/\sqrt{4\pi nm_i}$.

On small scales, the spectrum of turbulence is affected by damping.
Damping becomes important at a wave vector $k_c$ when the damping time
$\Gamma_{k}^{-1}$
becomes comparable or shorter than the  cascading time $\tau_{cas}$. Beyond this
wavevector the turbulence spectrum falls off rapidly.
In fully ionized plasma, there are basically two kinds of damping. Collisional
damping is important on scales greater than the Coulomb collision mean free path
$\lambda_{\rm Coul}$. In solar flares $\lambda_{\rm Coul}\sim 5\times 10^{7}{\rm
cm}\left(\frac{T}{10^7{\rm
K}}\right)^{2}\left(\frac{10^{10}{\rm cm}^{-3}}{n}\right)$ and the relevant
scales are shorter so that collisionless damping is dominant.
Taking into account interactions with thermal and nonthermal particles,
Petrosian, Yan \& Lazarian (2006, henceforth PYL06) studied damping of fast
modes in solar corona condition and showed that most the damping is also highly
anisotropic. For most angles of propagation the inertial range is truncated at
large scales. Only quasi-parallel and quasi-perpendicular waves reach short
scales comparable to ion gyroradius%
\footnote{The quasi-perpendicular most likely will be damped because of magnetic
field wanderings (see PYL06).}. 
For plasma $\beta_p=P_{gas}/P_{mag}\lesssim 0.1$, the damping due to electrons dominates as it's easier for electrons to 
catch up with the waves. The corresponding damping rate is  (Ginzburg 1961, YL02)
\begin{eqnarray}
\Gamma_{L} & = & \frac{\sqrt{\pi\alpha\beta_p}}{2}\omega\frac{\sin^{2}\theta}{\cos\theta}\exp\left(-\frac{\alpha}{\beta_p\cos^2\theta}\right),
\label{Ginz}
\end{eqnarray}
where $\alpha=m_e/m_H$, $\theta$ is the angle between the wave vector and the magnetic field. By equating the above equation and eq.(\ref{tcasfast}), we obtain
 the cutoff scale of turbulence owing to the collisionless damping,
\be
k_c L=\frac{4M_A^4\cos^2\theta}{\pi\alpha\beta_p\sin^4\theta}\exp\left(\frac{2\alpha}
{\beta_p\cos^2\theta}\right).
\label{landauk}
\ee
In what follows we shall use this spectrum of
fast modes to calculate the acceleration and confinement of particles by fast modes in solar
flare conditions.

\begin{table}
\caption{Notations in this paper.}
\begin{tabular}{|l l|}
\hline
a & power law index of nonthermal particle distribution\\
A(E) & acceleration rate\\
$E_0$ & lower energy limit of nonthermal particles\\
{\bf k}& wave vector\\
$\omega$ & wave frequency\\

$k_c$& turbulence cutoff wavenumber due to damping\\
L& energy injection scale\\
$\delta V$& injection speed\\
$M_A$ & Alfv\'enic Mach number $\delta V/v_A$\\
n & number density of the corona\\
N(E) & number density of nonthermal particles per energy bin\\
$N_0$ & N(E) at the lower energy limit $E_0$\\
T & corona temperature\\
$v_A$ & Alfv\'en speed\\
v & particle speed\\
$\mu$ & cosine of particle pitch angle\\
W({\bf k})& turbulence spectral energy density\\

$\alpha$& ratio of electron mass $m_e$ to ion mass $m_i$\\
B& magnetic field\\
$\beta_p$& ratio of gas pressure to magnetic pressure\\
$\beta$& v/c\\
$\beta_A$& $v_A/c$\\
$\Gamma$ & wave damping rate\\
$\tau_{cas}$ & cascading time of the turbulence\\
$\tau_{acc}$ & acceleration time scale\\
$\tau_{loss}$ & energy loss time scale of particles\\
$\tau_{esp}$ & escaping time of CRs from the system\\
$\eta$ &$\cos\theta$\\
$\eta_c$ &$v_A/v$\\
$\lambda_{Coul}$& Coulomb collisional mean free path\\
$\lambda_\|$ &parallel mean free path of CRs\\
$\theta$& pitch angle of a fast modes\\
$\delta \theta$& variation of $\theta$ in turbulence\\
\hline
\end{tabular}
\end{table}

\begin{table}
\caption{The physical parameters of solar flares we adopted.}
\begin{tabular}{|c|c|c|c|c|}
\hline
T(KeV)&n(cm$^{-3}$)&$\beta_p$&L(cm)&$M_A$
\\
\hline
1&$10^{10}$&$0.01, 0.04, 0.1$&$10^9$&$0.3$\\
\hline
\end{tabular}
\end{table}

In view of various difficulties with quasilinear theory (QLT), a number of
nonlinear theories (NLT) have been proposed (see Dupree 1966; V\"olk 1973;
Jones, Kaiser
\& Birmingham 1973; Goldstein 1976; Felice \& Kulsrud 2001; Matthaeus et al. 2003; Shalchi 2005).
Based on particle trapping due to large scale magnetic perturbations (V\"olk
1975), we developed a nonlinear formalism in YL08 to treat cosmic ray scattering
in MHD turbulence. In view of these progresses, we believe the time is ripe to
investigate the stochastic acceleration by the tested model of MHD turbulence in
solar flares. In  \S 2 we describe how the nonlinear effects are formulated. In \S 3 and \S 4 we 
present results on acceleration and confinement of the particles, and in \S 5 and \S 6 we present a
brief discussion and summary of the results.  Some mathematical details are given in the appendix.

\section{Nonlinear theory for particle acceleration}

The interaction of charged particle with MHD turbulence has been mostly described by the 
quasilinear theory (QLT). While the QLT allows easily to treat the CR dynamics in a local magnetic
field system of reference, a key assumption in the QLT, that the particle's orbit is unperturbed,
makes one wonder about the limitations of the approximation. 
While QLT provides simple physical insights into particle interaction with turbulence, it is known
to be an approximation. Recently there has been a surge
of interest in extending the treatment of this problem beyond the QLT. Examples of this are recently 
developed 
nonlinear guiding center theory by Matthaeus et al. (2003), weakly nonlinear theory by Shalchi et al. 
(2004) and second-order 
quasilinear theory by Shalchi (2005a). Most of the analysis so far are confined to slab and 2D 
model of MHD turbulence. In YL08, we extended the nonlinear treatment to models of large scale 
turbulence and obtained reasonable results for the diffusion coefficients. Here, we adopt the same 
approach for treating the particle acceleration for conditions appropriate for the solar corona.

In MHD turbulence, there are basically two types of interactions: gyroresonance and transit time 
acceleration (TTD). The resonant condition, for a particle of velocity $v$, pitch angle cosine $\mu$ 
and Lorentz factor $\gamma$, is $\omega-k_{\parallel}v\mu\approx n\Omega/\gamma$ ($n=0, \pm 1,2...$),
where $\omega$ is the wave frequency, $k_\|$ is the projection of wave vector along the local 
magnetic field $B$ and $\Omega=eB/(mc)$
is the gyro-frequency of  the particle (charge $e$ and mass $m$). 

TTD formally corresponds to 
$n=0$ and it requires compressible perturbations. When particles are trapped by moving in the 
same speed with waves, an appreciable amount of interactions can occur between waves and 
particles. Since head-on collisions are more frequent than that trailing collisions, particles gain 
energies. For small amplitude waves, particles should have parallel speed comparable to wave 
phase speed ($v_{ph}\approx \omega/k$) to be trapped in the moving mirrors. This gives rise to the above 
Cherenkov condition. In general, the momentum diffusion coefficient due to interactions (including both gyroresonance and TTD)
 with compressible modes can be written as (Yan \& Lazarian 2004)
\bea
D_{pp}&=&\pi\Omega m^2v_A^2(1-\mu^{2})\nonumber\\
&&\int_{\bf k_{\rm min}}^{\bf k_c}d^3k
\frac{W_\bot({\bf k})}{U_B}R_n(k_{\parallel}v_{\parallel}-\omega\pm n\Omega) J^{'2}_n\left(u\right),
\label{genrDpp}
\eea
where  $u=k_\perp v_\perp/\Omega$ and $U_B=B^2/8\pi$ 
is the energy density of mean magnetic field, $W_\bot({\bf k})$ is the kinetic energy of the turbulence motions perpendicular to the magnetic field. For 
fast modes in low $\beta$ medium, the velocity perturbations are perpendicular to the magnetic field (Yan, Lazarian \& Draine 2004), thus $W_\bot({\bf k})=W({\bf k})$ (see Eq.\ref{fastspec}). In quasi-linear approximation, particle's orbit is unperturbed and therefore the above resonance 
condition should be strictly observed so that $R_n(k_{\parallel}v_{\parallel}-\omega\pm n\Omega)=\delta(k_\|v \mu-\omega\pm n\Omega)$. 

However, the assumption of unperturbed orbit is not exact, as shown in YL08, the large scale motion (particularly compression) changes particles' pitch angle due to the 
conservation of adiabatic invariant. 
Because of the perturbation (especially compression) of magnetic field along the particle's 
trajectory, pitch angle is changing according to 
$\Delta v_\|/v_\bot\simeq (<\delta B^2_\|>/B^2)^{1/4}$, where $v_\|$ and $v_\bot$ are the particle 
speed along and perpendicular to the magnetic field,  and $\delta B_\|$ is the large scale 
perturbation of magnetic strength along the mean field. In this case, the 
resonance function will be broadened (YL08):
\be
R_n(k_{\parallel}v_{\parallel}-\omega\pm n\Omega)=\frac{\sqrt{\pi}}{\pi|k_\|\Delta v_\||}
\exp\left[-\frac{(k_\|v \mu-\omega\pm n\Omega/\gamma)^2}{k_\|^2\Delta v_\|^2}\right]
\label{Rn}
\ee  
This replaces the $\delta$ function in the QLT formula both for TTD ($n=0$) and gyroresonance interactions ($n=\pm 1$). This modification is particularly important for TTD because of the existence of a
critical pitch angle cosine  $\mu_c\sim v_A/v$ given by the
Cherenkov resonance condition. Physically, TTD take places on all scales while gyroresonance happens on a local scale. Therefore the large scale trapping is more influential to TTD. 

From this we can the obtain the systematic acceleration rate (see Petrosian \& Liu 2004)
\be
A(E)=\frac{\partial[vp^2D(p)]}{4p^2\partial p}, 
\label{ae}
\ee
where for isotropic wave distribution, 
\be
D(p)=\frac{1}{2}\int_{-1}^1 D_{pp}d\mu
\label{dp}
\ee
is pitch-angle averaged coefficient. The acceleration time is then given by
\be
\tau_{acc}=E/A(E)
\label{tauacc} 
\ee

\section{Particle Acceleration by Fast Modes}

Gyroresonance requires electromagnetic perturbations at a particular scale, e.g. 
$k_{res}\sim \Omega/v_\parallel\sim r_g^{-1}$. Given the parameters we adopt here, 
$k_{res}L>10^{9}$ for non-relativistic electrons. This is certainly beyond MHD regime 
Indeed nonthermal protons can be 
accelerated through gyroresonance with the fast modes. Inserting  the spectrum of fast modes from equations (\ref{fastspec}) into equation(\ref{genrDpp}), 
we get 
\bea
D^G_{pp}&=&\frac{v\sqrt{\pi}M_A^2m^2v_A^2(1-\mu^2)}{2LR^2}\int_{1}^{k_{max}L}dx\int_0^1 d\eta
\frac{ x^{-\frac{5}{2}}}{\eta \Delta \mu}[J_1(w)']^2\nonumber\\
&&\exp\left[-\frac{(\mu-\frac{v_A}{v}\pm\frac{1}{x\eta R})^2}{\Delta \mu^2}\right],
\label{fastgyro}
\eea
Nevertheless, our calculations shows the overall contribution from gyroresonance is much smaller 
than that from TTD. In fact for the adopted parameters, there is no acceleration from gyroresonance 
for protons with energies less then $<$4MeV because of the damping of fast modes. For higher 
energy protons, acceleration rate from gyroresonance is 2-3 orders of magnitude lower than that from TTD.

On the other hand, TTD interactions can happen on all scales or wavevectors $k$ and appears to be 
the dominant acceleration mechanism for protons and the only channel to direct the energy from 
MHD turbulence to electrons. The corresponding momentum diffusion, in the NLT limit, 
can be obtained in a similar way, 

\bea
D_{pp}&=&\frac{\sqrt{\pi}M_A^2m^2v_A^2v}{2LR^2}(1-\mu^2)\int_0^1 d\eta
\int_{1}^{k_cL}\frac{J_1^2(u)}{\Delta \mu}x^{-\frac{5}{2}}dx\nonumber\\
&&\exp\left[-\frac{[\mu-v_A/(v\eta)]^2}{\eta \Delta \mu^2}\right],
\label{fastTTD}
\eea
where $\Delta \mu=\Delta v_\|/v$, $R=vk\gamma/\Omega$ is the dimensionless rigidity, $M_A$ is 
the Alfv\'enic Mach number, and $\eta=\cos\theta$. 
Combining equation (\ref{fastTTD}) with equation (\ref{ae},\ref{dp}) we get the acceleration rate
\begin{eqnarray}
A(E)&=&\frac{\sqrt{\pi}M_A^{\frac{3}{2}}p v_A^2}{8LR^2} \times \int_0^1 d\mu \sqrt{1-\mu^2} 
 \int^1_0 \eta^{-1} d\eta \int_1^{k_cL} x^{-\frac{5}{2}} d x \nonumber\\
&&\left\{2\left[1-\frac{v_A}{v\eta}\frac{(\mu-v_A/v\eta)}{\Delta \mu^2}\right]J_1^2
\left(u\right) +u J_1\left(u\right)\left[J_0\left(u\right)\right.\right.\nonumber\\
&-&\left.\left.J_2\left(u\right)\right] \right\}\nonumber\\
&&\exp\left[-\frac{(\mu-v_A/v\eta)^2}{\Delta \mu^2}\right].
\label{NLT}
\end{eqnarray}


{\it Comparison of NLT with QLT:}
The QLT momentum diffusion and direct acceleration rates due to TTD are obtained by adopting the delta function as the resonance function $R_n(k_\|v_\|-\omega)=\delta(k_\|v_\|-\omega)$:
\begin{eqnarray}
D_{pp}&=&\frac{\pi M_A^2m^2v_A^2v}{2LR^2}(1-\mu^2)\int_0^1 \eta^{-1} d\eta\int_{1}^{k_cL}x^{-\frac{5}{2}}dx
J_1^2(u)\nonumber\\
&&\delta\left(\mu-\frac{v_A}{v\eta}\right)
\label{dpp}
\end{eqnarray}
 
and

\begin{eqnarray}
&&A(E)=\frac{\pi M_A^2 pv_A^2}{8LR^2}\left(1-\frac{v^2_A}{v^2\eta^2}\right)\times \int_0^1 \eta^{-1} d\eta\int_{1}^{k_cL}x^{-\frac{5}{2}}dx\nonumber\\
&&\left\{2 J_1^2\left(u\right) +u J_1\left(u\right)\left[J_0\left(u\right)-J_2\left(u\right)\right]\right\}.
\label{QLT}
\end{eqnarray}

\begin{figure}
\plotone{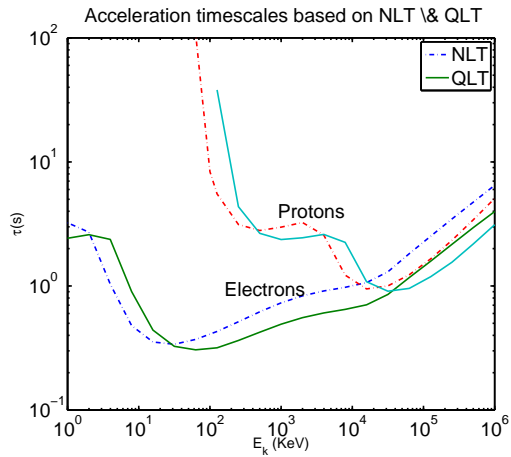}
\caption{The acceleration times based on NLT (dashdot line, Eq.\ref{NLT}) and QLT (solid line, Eq.\ref{QLT}). The lower thick lines represent the result for electrons and the upper thin lines are for protons.}
\label{qltnlt}
\end{figure}

For electrons and low energy ions ($\lesssim 100$KeV), in magnetic field of $B\sim 10^2$ G,
 $v_\perp/\Omega_e < c/\Omega_e \simeq 14{\rm cm}$, so for most
 angles $k_\perp v_\perp/\Omega<<1$ and the Bessel
 function can be replaced by
 the first order approximation $J_{n}(x)\simeq(x/2)^{n}/n!$. Thus inserting $k_c$ given by eq.(\ref{landauk})
into equation (\ref{QLT}), 
we obtain for the QLT limit
\begin{eqnarray}
A(E)&=&\frac{\pi pv_A^2M_A^2}{4L}\left\{M_A^2\left(\frac{2}{\beta_p}-\frac{\eta_c^2}{\alpha}\right)\left[\Phi_I\left(\frac{1}{\eta_c}\sqrt{\frac{\alpha}{\beta_p}}\right)\right.\right.\nonumber\\
&-&\left.\Phi_I\left(\sqrt{\frac{\alpha}{\beta_p}}\right)\right]+(1-\eta_c^2)+(1+\eta_c^2)\log\eta_c\nonumber\\
&+&\left.\frac{2M_A^2}{\sqrt{\pi\alpha \beta_p}}\left[\exp\left(\frac{\alpha}{\beta_p}\right)-\eta_c\exp\left(\frac{\alpha}{\beta_p\eta_c^2}\right)\right]\right\},
\label{errfunc}
\end{eqnarray}
where $\Phi_I$ is the imaginary error function. 
The NLT and 
QLT result  are compared in Fig.\ref{qltnlt}. We see that the two results qualitatively agree with each other
so that for most cases one can use  the acceleration rate using QLT. 

\begin{figure}
\plotone{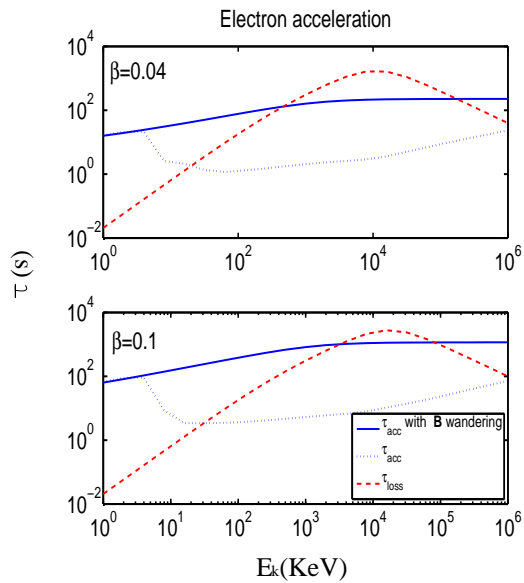}
\caption{The influence of field line wandering to TTD acceleration by fast modes. 
The solid and dotted lines are the averaged accelerate times
taking into account field line wandering; the dotted lines
are the results without accounting for field line wandering. The acceleration rate is decreased due to averaging because field line wandering 
increases damping at large angles.}
\label{eacc1}
\end{figure}

\begin{figure}
\plotone{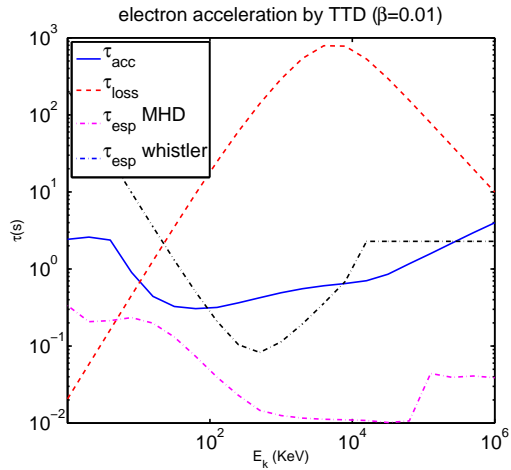}
\caption{times involved in the acceleration of electrons. The solid line refers to the acceleration time. The dashed line represents energy loss time due to Coulomb collisions and synchrotron radiation. The dashdot lines give the escaping times. Scattering with MHD modes (lower curve) is ineffective;   the actual confinement is dominated by the interaction with the quasi-parallel whistler modes (upper curve). The cross of accelerate time and energy loss time determines the threshold at the lower energy end. The acceleration continues till the cross at the higher energy end.}
\label{eacc2}
\end{figure}

\begin{figure}
\plotone{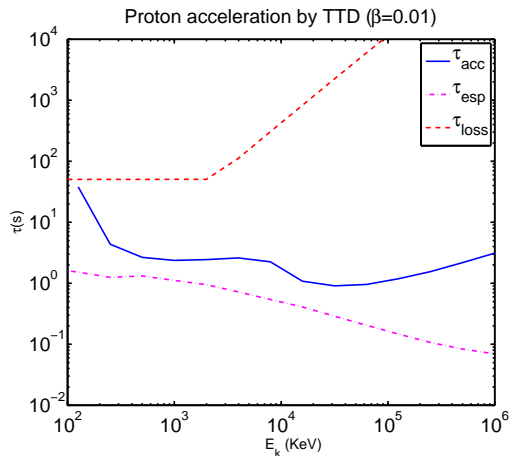}
\caption{Same as Fig.\ref{eacc2}, but for protons. The acceleration starts with super-Alfv\'enic particles ($\sim 2(T/10^7K)/\beta_p$KeV). The confinement is owing to the scattering by the MHD modes (dashdot line). The acceleration continues to $0.1$GeV, beyond which the escaping becomes much faster than the acceleration.}
\label{pacc}
\end{figure}

\emph{Effects of field line wandering:} Unless the medium is strongly magnetized ($\beta_p\lesssim 0.1$), 
field line wandering can not be neglected (YL04). As shown in PYL06, field line wander about $\lesssim 15^\circ$ around $90^\circ$. Since for high velocity particles, $\eta\approx v_A/(v\mu)\ll 1$ and the quasi-perpendicular modes dominate. Thus this angle variation around $90^\circ$ must be taken into account. The damping rate should be averaged over the range $90^\circ-\delta\theta \thicksim 90^\circ$. Equating cascading rate (see Eq.\ref{tcasfast}) and the damping rate (Eq.\ref{Ginz}) averaged over $\delta\theta$ (PYL06)
\bea
\tau_{cas}^{-1}&=& \frac{\sqrt{\alpha\pi\beta_p}}{2\delta\theta}kv_A\int^{\delta\theta}_{0}\frac{\exp(-\alpha/\beta_p \phi^2)}{\phi}  d\phi\nonumber\\
&=&\frac{\sqrt{\alpha\pi\beta_p}}{4\delta\theta}kv_AE_1\left(\frac{\alpha}{\beta_p \delta\theta^2}\right),
\label{kaverage}
\eea
and combining it with the variation of angle $\theta$ due to field line wandering $\delta\theta\simeq (M_A^2/kL)^{1/6}$, one gets the averaged damping wavenumber ${\bar k_c}$. For $M_A=0.3$, it has only solution for $\beta_p>0.04$. For $\beta_p=0.1$, $\bar{k_c}L\sim 16$. In Fig.\ref{eacc1}, we compare the acceleration rate with and without field line wandering. As can be seen, the acceleration is decreased in the case of $\beta_p>0.04$. This can be explained as follows. According to the Landau resonant condition $k_\|v_\|\simeq\omega$, the higher the energy, the closer to $90^\circ$ the resonant wave vector. As shown in Eq.(\ref{kaverage}), the rapid increase of the cutoff wavenumber near $90^\circ$ is smeared out due to the field line wandering. Acceleration rate is accordingly modulated for particles with different energies. This result shows that field line wandering has a noticeable impact on the acceleration and must be accounted for. Because of the field line wandering, acceleration is substantially reduced when $\beta_p\gtrsim 0.04$. For $\beta_p\lesssim 0.01$, effects of field line wandering is negligible due to the dominance of magnetic energy. In figures \ref{eacc2},\ref{pacc}, we compare the acceleration time due to TTD interaction with other times for electrons and protons. We see an increase of acceleration rate at $\sim 10$KeV for electrons and $\sim 10$MeV for protons because of field line wandering. The acceleration times drop suddenly because the wave damping drops dramatically when the corresponding $\theta$ approaches $90^\circ$. The fact that the acceleration rate increases substantially at $\beta_p< 0.04$ sets a critical value for the plasma $\beta_p<\beta_{cr}$ provided that the stochastic acceleration by MHD turbulence is the dominant mechanism for generating high energy particles. Here $\beta_{cr}$ is approximately determined by $M_A$ through the Eq.(\ref{kaverage}).
  
\section{Confinement of Particles}

For efficient acceleration particles should be confined within the system for a sufficient duration. 
This is particulary nevessary for TTD where particles gain energy only in the parallel direction. 
They need to be isotropized in order to reach high energies. 
The isotropization is dominated by Coulomb collisions for low energy protons ($\lesssim100$ keV). 
The Coulomb scattering rate drops quickly with the increase of energy. For higher energy particles, 
the scattering is owing to the interaction with turbulence. The same turbulence that 
accelerate the particles statistically also cause pitch angle scattering and thereby provide 
confinement for these particles. 

We showed in YL08 that quasilinear approximation is not valid for scattering from TTD and 
resonance broadening due to large scale trapping must be taken into account. This removes the divergence at the $90^\circ$ and leads to a finite mean free path,   
\be
\lambda_\|=\frac{3}{4}\int^1_0 d\mu \frac{v(1-\mu^2)^2}{(D^T_{\mu\mu}+D^G_{\mu\mu})}
\ee
where the pitch angle coefficient due to TTD is
\bea
D^T_{\mu\mu}&=&\frac{v\sqrt{\pi}(1-\mu^2)}{2LR^2}\int_{1}^{k_{c}L}dx\int_0^1 d\eta
\frac{x^{-5/2}\eta}{\Delta \mu_\|}J_1^2\left(u\right)\nonumber\\
&&\left(\eta+\frac{\mu v_A}{v}\right)^2\exp\left[-\frac{2(\mu-v_A/v)^2}{\Delta \mu^2}\right].
\label{fastDmumu}
\eea

TTD dominates the scattering for most pitch angles. For high speed particles with small pitch angles, however, TTD interaction is negligible as indicated from the Chenrenkov condition and gyroresonance takes over. We showed in YL08 that QLT provides a good approximation for gyroresonance since it is a local interaction on small scales. We adopt here the QLT result for the pitch angle scattering due to gyroresonance,
\bea
D^G_{\mu\mu}&=&2\pi\Omega^{2}(1-\mu^{2})\int_{\bf k_{min}}^{\bf k_c}d^3k \frac{W({\bf k})}{U_B}\nonumber\\
& &\delta(k_{\parallel}v_{\parallel}-\omega\pm \Omega) \left(\eta+\frac{\mu v_A}{v}\right)^2 J_1'\left(u\right)^2.
\eea
The confinement time of the particles can be approximated by (see Petrosian \& liu 2004)

\be
\tau_{esp}=\frac{L}{\sqrt{2}v}\left(1+\frac{\sqrt{2}L}{\lambda_\|}\right)
\label{esptime}
\ee
This time scale is also shown in Figs.\ref{eacc2},\ref{pacc}.
The threshold for TTD interaction is $v_\|\gtrsim v_A$ as indicated by the Cherenkov resonance condition. That's why the confinement time for the protons $L^2/(\lambda v)$ rises around $E_k\simeq 100$KeV and drops on high energy end (see Fig.\ref{pacc}). Acceleration continues up to $\sim 10$MeV, beyond which the confinement time becomes much smaller than the acceleration time.

Indeed pitch angle varies with the variation of magnetic field due to conservation of adiabatic invariant. The large scale slow variation of magnetic field (compared to the gyration of particles), however, is inefficient in accelerating the particles (see Cho \& Lazarian 2006). Therefore, while for acceleration QLT and NLT estimates are comparable, scattering and spatial diffusion has to be studied with nonlinear approach to remove the $90^\circ$ degeneracy and to obtain finite mean free paths (see YL08).

For electrons, lack of gyroresonance (except for high energy electron with $r_{res}>k_c^{-1}$), the interaction with MHD modes is not enough to confine the particles. A cutoff of the spectrum $\gtrsim 100$keV would occur if there are no other interactions. Higher frequency whistler modes have been considered as one candidate to interact with electrons (Wentzel 1976; Melrose \& Brown 1976; Bespalov et al. 1991; Stepanov et al. 2007). However, due to their anisotropy (Cho \& Lazarian 2004), the interaction with whistler turbulence is very inefficient similar to the case of CR scattering by Alfv\'enic turbulence (Chandran 2000; YL02). On the other hand, parallel propagating whistler waves, can be generated by kinetic instabilities, e.g., beaming instability, anisotropy instability, cyclotron instability (Akhiezer et al. 1975; Tsytovich 1977; Neubert \& Banks 1992) and be an efficient agent to scatter electrons. Through the TTD acceleration, an anisotropic distribution of particle with respect to the magnetic field can be generated, which can induce cyclotron instability (Lazarian \& Beresnyak 2006). Leakage during solar flares is also possible to create beaming instability. Scattering by the whistler modes limits the bulk speed of the particles close to the phase speed of the whistler waves. The waves are generated under the cyclotron resonance $\omega=kv_\|-\Omega_e/\gamma_e$. Combined with the dispersion relation of whistler modes
\be
\omega=\frac{\Omega_e k^2}{k^2+(\omega_{pe}/c)^2},
\ee
from which we can get the resonant wave number $k_{res}(\gamma_e)$. The group velocity of the resonant modes is then  obtained by inserting it into $v_g(\gamma_e)=\partial \omega/\partial k$, which is a function of the electron energies. The confinement time will be approximately $L/v_g$. For high energy electrons ($E_k\gtrsim 10$MeV), the resonant wave modes get smaller than the gyrofrequency of thermal ions and the group velocity $v_g \sim v_A$ for the wave modes moving parallel to the magnetic field. Fig.\ref{eacc2} shows that the parallel propagating whistler modes are able to confine the electrons during the acceleration.

The whistler waves are subjected to both Coulomb damping and collisionless damping. For electrons with energies $<1$GeV, however, whistler wave can be excited and provide effective scattering for the electrons (Dorman 1996). The detailed study is beyond the scope of this paper.

\section{Discussion}

In this paper we have discussed the effects that the cascade of
turbulence has on acceleration and heating of Solar corona. There the energy is injected
at large scales much larger than any plasma scale concerned and this
justifies a magnetohydrodynamic treatment of the large
scale motions. Due to recent insights into the physics of MHD cascade
and its interaction with charged particles we reduced a complex
problem of acceleration and heating to a more manageable problems of interactions of
Alfv\'en, slow and fast modes with plasma and energetic particles.

Alfv\'enic turbulence is inefficient in scattering and accelerating particles because of its anisotropy (Chandran 2000, YL02). Fast modes, instead, have been identified as the MHD turbulence modes dominating the interaction with cosmic rays (YL02,YL04). In this paper, we apply the result to the stochastic acceleration in solar flares.

We assume that the MHD turbulence is strong, i.e. that the critical balanced condition is satisfied for Alfv\'enic modes. Therefore, to describe fast mode of MHD turbulence we appeal to the results by Cho \& Lazarian (2002, 2003) on the scaling and coupling of Alfv\'enic and fast modes. The corresponding papers claim the isotropy of the fast modes. One may expect to see deviations from isotropy, however. For instance, slab Alfv\'en modes created by streaming instabilities are subject to non-linear damping by the ambient Alfv\'enic turbulence (YL02, YL04, Farmer \& Goldreich 2004, Lazarian \& Beresnyak 2006). Beresnyak \& Lazarian (2008) showed that the corresponding damping of the quasi-parallel modes depends on the angle between their ${\bf k}$ vector and
the direction of magnetic field. One might expect quasi-slab fast mode to be subject to a similar damping by strong Alfv\'enic turbulence. This, however, has not been demonstrated so far. 

If at the injection scale $\delta B\ll B$, the Alfv\'enic turbulence is weak (see Galtier et al 2000) and develop a cascade with $k_{\|}=const$. The interaction of the weak Alfv\'enic turbulence with other modes can 
be very different from that of the strong Alfv\'enic turbulence. For instance, it was shown by Chandran (2005) that fast modes develop anisotropy (i.e. the energy in the quasi-slab modes is reduced) owing to their interaction with Alfv\'en modes in the weak  regime. However, such the Alfvenic weak turbulence has
a limited inertial range (see discussion in Cho, Lazarian \& Vishniac 2003) and at sufficiently large $k$ transfers into a strong turbulence. Moreover, magnetic reconnection in Solar Flares should produce perturbations $\delta B\sim B$, which should induce strong turbulence from the very beginning.

Strong Alfv\'enic turbulence is characterized by $k_{\bot}\gg k_{\|}$ with the GS95 relation $k_{\|}\sim k_{\bot}$ defining a cone in the Fourier space where most of the turbulent energy resides. However, the aforementioned relation should not be understood too literally. The energy outside the cone is not zero and the modes with $k_{\|}>k_{\bot}$ are present. Such Alfv\'enic modes are weakly interacting even being a part of the strong Alfv\'enic turbulence. Therefore, the Chandran (2005) model is applicable to them.  Nevertheless, according to Cho, Lazarian \& Vishniac (2002) the energy in these modes is
exponentially reduced\footnote{Because of this exponential reduction the
rates of Alfv\'enic scattering in YL02, which was the first study to take
into account the modes outside the GS95 cone, were still grossly
subdominant to the fast modes.}. Therefore, assuming that the Alfv\'enic
turbulence is injected at a scale much larger than the typical gyroradius
of the energetic particles, we did not consider the anisotropies,
introduced by the process of the fast wave cascading induced by Alfvenic
modes with large $k_{\|}$ and small $k_{\bot}$ (cf. Chandran 2005)."

The model for MHD turbulence that we adopted in the paper is the turbulence where the back-reaction of energetic particles is limited to changing the cut-off of the turbulence. A more fundamental modifications of turbulence are conceivable, however. For instance, Lazarian \& Beresnyak (2006) argued that compressions of the fluid of energetic particles may result in the gyroresonance instabilities that can induce an additional quasi-parallel component of the Alfvenic waves. If true, these waves would interact with fast modes in the manner described by Chandran (2005), which would affect the fast mode isotropy. However, if substantial portion of energy resides within these quasi-slab modes, their major effect will be the direct gyroresonance acceleration of energetic particles. We felt that the modification of MHD turbulence arising from the instabilities within energetic particles, e.g. by  the process in Lazarian \& Beresnyak (2006), is beyond the scope of the present paper.  

Apart from the issue of isotropy, there are potential issues related to the exact scaling of fast modes. One dimensional numerical simulations in Suzuki, Lazarian \& Beresnyak (2007) indicate that fast modes may develop a shock-like cascade, which differs from the finding in 3D MHD calculations in Cho \& Lazarian (2002, 2003). We adopted the model from the latter works, but wait for higher resolution 3D numerical runs to rectify the scaling.    

In addition to being strong, MHD turbulence that we considered was balanced in the sense that the
equal flux of energy was assumed in every possible direction. The properties
of imbalanced turbulence (see Cho, Lazarian
\& Vishniac 2002; Lithwick, Goldreich \& Sridhar 2007; Beresnyak \& Lazarian 2007, Chandran 2008) can be very different from the balanced one. However, we expect the flow of Alfv\'en waves, which constitute the weak turbulence, to be  subjected to reflection within a Solar corona environment that we
deal with. As the result we expect only marginal imbalance for the problem that we deal with. 

A threshold for the TTD acceleration is $v\gtrsim v_A$ set by the Cherenkov resonance condition. In the low $\beta_p$ environment, the thermal protons can not be accelerated. The low energy threshold would be $\sim 2(T/10^7)/\beta_p$KeV. Thermal electrons, instead, can have TTD interactions unless the plasma beta is too low $\beta_p\lesssim m_e/m_p$. Our calculations show that TTD acceleration dominates over gyroresonance for the energy range we consider. The acceleration efficiency decreases with the plasma $\beta_p$ as damping of the fast modes increase with $\beta_p$.

Particle acceleration rate depends on the wave spectrum and the wave damping rate are
partially determined by the particle spectrum. 
In general, it is required that a self-consistent treatment of the evolution of turbulence and 
particle acceleration. The calculations above assumes that the spectrum of the turbulence is determined 
by the cascade and damping by thermal particles. 
Our numerical calculation of these integrals shows that up to $\sim$ 10\% of 
total energy of turbulence is being transferred to nonthermal particles for the given set of parameters.
In most cases, the back reaction of nonthermal particles is negligible as the damping due to the 
interaction with these nonthermal particles is smaller than thermal damping taking into account 
field line wandering. In some large flares, however, a large amount of particles need to be accelerated 
from the thermal reservoir to energies $\gg k_BT$ (PYL06). In this case, the damping by nonthermal particles
can be significant and one needs to solve the coupled equation of the evolution of turbulence and
particles simultaneously. 

Isotropization is important for the process of acceleration.
If there is not enough isotropization, the acceleration by TTD will stop quickly as only parallel velocity is increased during the process. Moreover, without enough scattering, particles will leave the system before they get accelerated. Scattering by fast modes is shown to be adequate for isotropization of protons. Different from the acceleration, the scattering by gyroresonance, even smaller than the TTD scattering rate, plays an essential role in determining the scattering of parallel moving particles, which can be a substantial portion of the particles because of the TTD acceleration.

For electrons, due to their small gyro-radii, gyroresonance does not occur with MHD turbulence except for those high energy electrons.
The actual scattering can happen with the plasma turbulence. The work by Cho \& Lazarian (2004) shows that whistler modes are even more anisotropic than Alfv\'en modes. We know gyroresonance is very inefficient with anisotropic turbulence.  Analogous to the MHD regime, parallel propagating modes may be generated by kinetic instabilities (see Tsytovich 1977) and be a candidate for interaction with electrons. The TTD interaction itself may induce gyroresonance instability through creating an anisotropic distribution of particles with respect to the magnetic field (Lazarian \& Beresnyak 2006). Our estimate shows that the whistler waves (for moderate energy) and Alfv\'en wave (for high energy) generated by the anisotropy instability can provide effective scattering and isotropization for the electron acceleration. 

Acceleration of particles by fast modes were previously studied by a number of authors, including Miller, Larosa \& Moore (1996), Schlickeiser \& Miller (1998). In their studies, turbulence is assumed isotropic with either Kolmogorov or Kraichnan spectrum. Although coupled equations of wave and particle evolutions were solved in Miller, Larosa \& Moore (1996), we feel that the one dimensional treatment they adopt is problematic, as the damping caused by the TTD interaction with particles is anisotropic. In this paper, we start with a more physically motivated and numerically tested of turbulence. We deal with fast modes, which are subject to much
more efficient linear dissipation.  On the small scales, however, because of the dissipation above, fast modes develop anisotropy. In our treatment, this anisotropy is strongly affected by field line wandering which is determined by the Alfv\'enic Mach number and plasma $\beta$ and the efficiency of the acceleration is substantially influenced accordingly. In addition, scattering and confinement are treated using the nonlinear theory we have recently developed (Yan \& Lazarian 2008). We believe that the approach that we developed is applicable beyond pure Solar Flare problems, e.g. it can be modified to study turbulent acceleration in the medium
within clusters of galaxies (see Brunetti \& Lazarian 2007).

\section{Summary}
We discuss stochastic acceleration in solar flares. The dominance of thermal damping  makes it possible to decouple acceleration and turbulence evolution. This simplifies the problem substantially so that we don't have to rely on simulation to resolve the problem. Our study can be summarized as below:
\begin{itemize}
\item Fast modes are sufficient to accelerate super-Alfv\'enic particles to high energies. Electrons can reach GeV energies through the process.

\item Both nonlinear and quasilinear approaches show TTD dominates the acceleration by MHD modes. For the acceleration, QLT result qualitatively agrees with the nonlinear result and appears to be a good approximation.

\item Confinement of protons can be realized through both gyroresonance and TTD with fast modes. Electrons, however, needs plasma perturbations to be confined.

\item Magnetic field line wandering reduces the acceleration efficiency in weakly ionized plasma and this sets an upper limit of plasma $\beta_p$ for solar flares. 
\end{itemize}

\begin{acknowledgments}
HY is supported by CITA and the National Science and Engineering Research Council of Canada. AL acknowledges
the support by the NSF ATM-0648699 and the NSF Center for Magnetic Self-Organization in Laboratory and Astrophysical Plasmas. 
VP acknowledges the support of NASA grant NNG05GAI90G and NSF grant ATM-0648750.

\end{acknowledgments}

\appendix

\section{energy loss rate}

In the paper, the acceleration time $\tau_{acc}=E/A(E)$ is compared with the energy loss time:
\bea
\tau_{loss}=E/\dot{E}_{loss}=\frac{\gamma-1}{4r_0^2(\pi n_ec\ln\Lambda/\beta+B_0^2\beta^2\gamma^2/9/m_ec)}
\eea 
where $r_0=2.8\times 10^{13}$cm is the classical electron radius and $\ln\Lambda=20$ in our regime (see Sturrock 1994). The ion loss is mainly due to Coulomb collisions with electrons and protons (Post 1956; Ginzburg \& Syrovatskii 1964). The loss due to electron-ion collisions is (Petrosian \& Liu 2004)

\bea
\tau_{loss}=\frac{\gamma-1}{2\pi r_0^2\alpha cn_e\left(\frac{q_i}{e}\right)^2}\cases{\frac{\sqrt{2}}{4}\frac{\beta_{te}}{\beta^2c^2}\ln^{-1} \Lambda\left(\pi/3\right)^{1/2} & for $\beta<\beta_{te}$\cr
\beta\ln^{-1}\left(\frac{m_e^2\beta^4c^2}{\pi r_0n_e\hbar^2}\right) & for $\beta_{te}\ll 1$\cr
\ln^{-1}\left(\frac{m_e^2c^2\gamma^2}{2\pi r_0n_e\hbar^2}\right) & for $1\ll \gamma\ll \frac{m_i}{m_e}$\cr
\ln^{-1}\left(\frac{m_em_ic^2\gamma^2}{4\pi r_0n_e\hbar^2}\right) & for $\frac{m_i}{m_e}\ll \gamma$\cr}
\eea

At low energies, the loss is dominated by ion-ion collisions,

\be
\tau_{loss}=\frac{\gamma-1}{4\pi r_0^2\alpha^2 c/\beta n_p(q_i/e)^2\ln \Lambda}
\ee

\end{document}